# Integrating lpGBT links into the Common Readout Units (CRU) of the ALICE Experiment

**E. Dávid**[a] **and T. Kiss**[a] on behalf of the ALICE Collaboration

[a] *HUN-REN Wigner Research Centre for Physics,*
*29-33 Konkoly-Thege Miklós Str, H-1121 Budapest, Hungary*
*E-mail*: erno.david@cern.ch, tivadar.kiss@cern.ch

ABSTRACT: In the ALICE read-out and trigger system, the present GBT and CRU based solution will also serve for Run4 without major modifications. By now, the GBT protocol has been superseded by lpGBT. Extensions of the ALICE system (e.g. the planned FoCal and ITS3 detector) will therefore require to use lpGBT while keeping the compatibility with the existing system. In this paper we show the implementation and testing of a possible integration of the lpGBT-FPGA IP into the CRU firmware, allowing the extension of the present system, keeping it more versatile and future-proof.

## Contents

## 1. Introduction

During the LS2 upgrade of the ALICE experiment a new DAQ and trigger system had been developed for Run 3 and Run 4. The sub-detectors are now connected to the ALICE DAQ and Trigger systems with radiation hard GBT [1] links through Common Read-out Units (CRU) [2] [3]. The GBT links enable the data taking and the delivery of timing and control with deterministic latency through a single optical cable connection.

By now, the GBT protocol has been superseded by lpGBT. Extensions of the ALICE system (e.g. the planned FoCal and ITS3 detector) will therefore require the use of lpGBT while keeping the compatibility with the existing system.

In this paper, we show the implementation and testing of a possible integration of the lpGBT FPGA IP into the CRU firmware. It allows the extension of the present system, keeping it more versatile and future-proof.

## 2. lpGBT and GBT Differences

Both the GBT and the lpGBT link protocols are implemented in rad-hard ASICs on the detector front-ends and as soft IPs that can be implemented in different FPGAs in the back-ends. The main differences between the lpGBT and the GBT link protocols can be summarized as follows.

Downlink (CRU to detector direction): The new lpGBT downlink speed is 2.56 Gb/s which enables the transmission of 64 bits in each 40 MHz LHC clock cycle. It contains 32-bit data and 4 control bits as the payload. This compares to the 4.8 Gb/s line speed of the GBT link which can transfer 120 bits in each LHC clock cycle which contains 80-bit data and 4 control bits. From the CRU user point of view, the different payload size means that the lpGBT downlink can transmit fewer trigger bits (32-bit instead of the previous 80-bit) to the Front End Electronics (FEE) in each cycle. The other important technical difference is that the lpGBT downlink requires a 320 MHz reference clock for the transmission compared to the 240 MHz reference clock for the GBT downlink.

Uplink (detector to CRU direction): The lpGBT supports two uplink speeds, 5.12 or 10.24 Gb/s, which means the transmission of 128 or 256 bits of data per LHC clock cycle. From the CRU user point of view, the 5.12 Gb/s speed version is nearly identical to the 4.8 Gb/s GBT transmission, while the 10.24 Gb/s link speed doubles the incoming data rate. The clock frequency of the lpGBT receiver interface has also changed to 320 MHz from the previous 240 MHz of the GBT RX.

## 3. CRU Firmware

The overview of the CRU firmware can be seen in Figure 1. In the CRU, the trigger and timing information is received by the TTC firmware module over a 10 Gbps Passive Optical Network (10G PON) link from an ALICE Local Trigger Unit (LTU), and passed down via the GBT transmit interface to the detector FEE. In the uplink path the detector data is received by the GBT receiver interface and either processed by the default packet processing modules or by a user-specific logic (depending on the type of the subdetector the CRU is connected to). Then it is transferred by the PCIe DMA engine to the RAM of the host computer (FLP servers) for further processing in the read-out software chain.

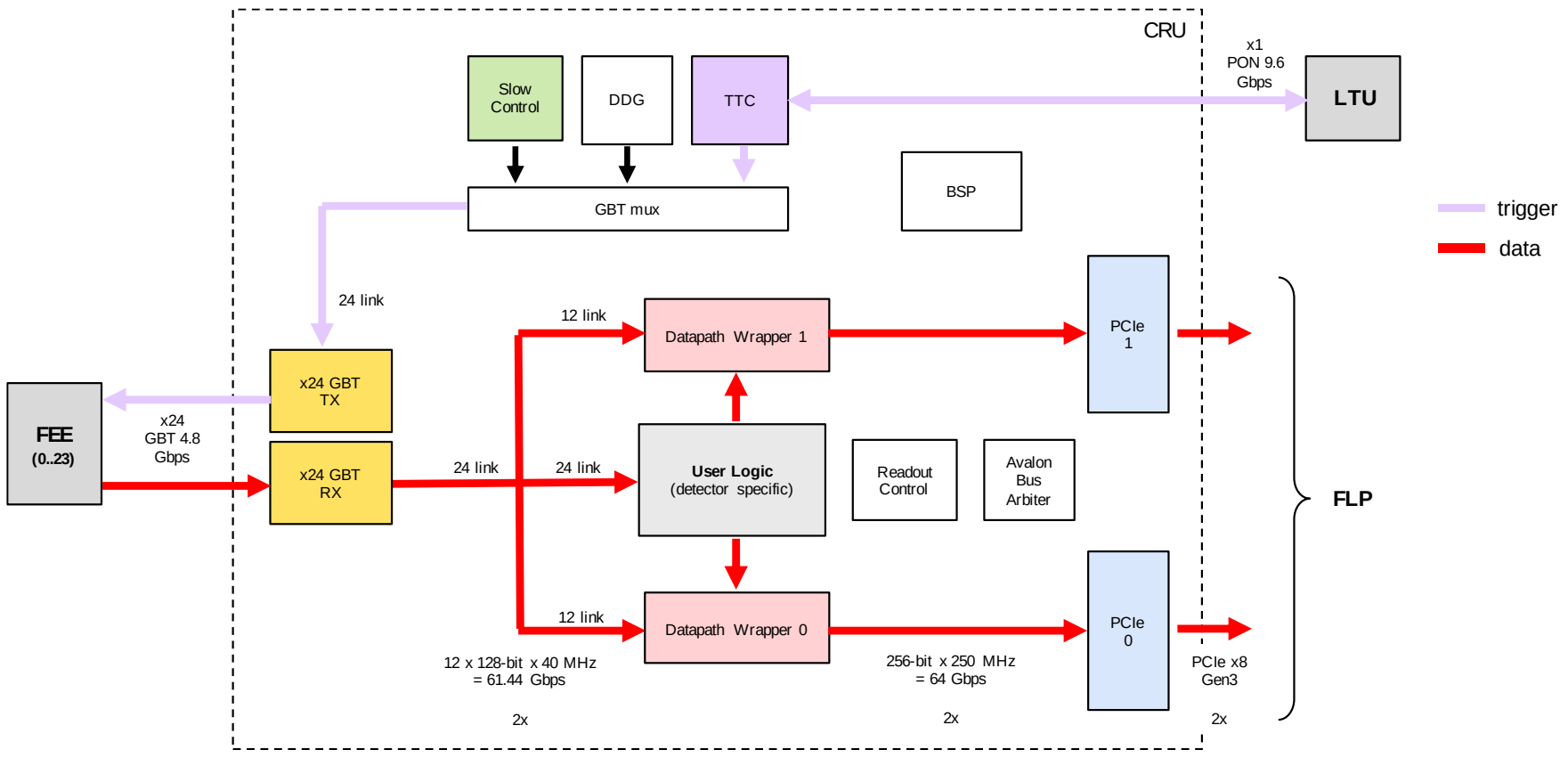

**Figure 1.** Overview of the CRU firmware.

## 4. lpGBT Integration

Figure 2 shows the current clock and trigger distribution within the CRU [4]. The 10G PON receiver extracts the trigger data bits and a 240 MHz receive clock and a valid bit that marks the 40 MHz LHC clock edges in this 240 MHz clock domain. In the GBT case, luckily the GBT link transmitters also use 240 MHz as their reference clock, so the clock recovered from the PON link needs a jitter cleaning only and can then directly be used as the GBT links reference clock. (The recovered clock is cleaned and distributed to the GBT transceiver banks by two external jitter cleaner Phase-Locked Loops (PLLs)) Thus, in the GBT case, the reception and further transmission of the trigger data just occur in the same clock domain and no further synchronization is needed.

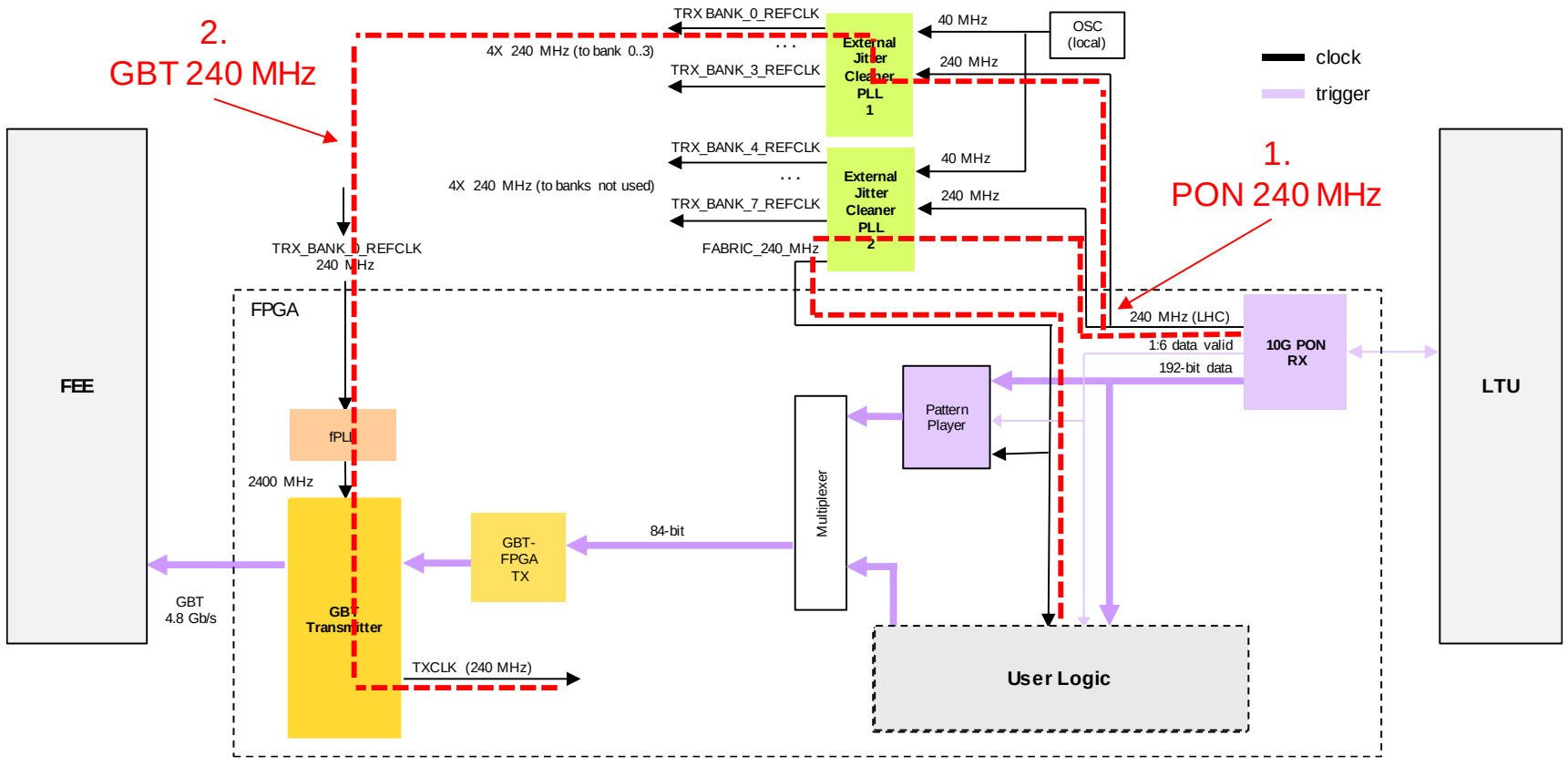

**Figure 2.** Overview of the GBT based clock and trigger distribution in the CRU firmware

However, the new lpGBT FPGA IP requires 320 MHz as the link transmitters reference clock. The new clock ratio is 4:3 which breaks the solution used by the current firmware. First, the 320 MHz lpGBT link clocks must be generated from the incoming 240 MHz clock synchronously, then the trigger data must be transmitted from the 10G PON receive clock domain to the lpGBT transmit clock domain strictly maintaining the phase alignment to the PON valid bit (marking the 40 MHz LHC clock cycles), in order to provide a deterministic latency through the CRU FPGA firmware.

To achieve this performance, the following solution has been developed. (Figure 3 summarizes the steps):

1. Divide the incoming 240 MHz PON RX clock by six with an internal I/O PLL (IOPLL) to generate a 40 MHz internal clock. Due to the frequency division, this 40 MHz internal clock can have 6 random phase positions relative to that edge of the 240 MHz clock which is marked by the PON valid bit (i.e. the LHC clock).
2. Therefore, this internal clock must be aligned to the PON valid bit to maintain its deterministic phase position. This is achieved by sampling the PON valid bit by the generated clock. If there is no match, a control state machine resets the PLL, and the procedure restarts.
3. With the same IOPLL we generate a 320 MHz clock too, which is in sync with the 40 MHz clock with a frequency ratio 8:1. The generated 320 MHz clock is then sent out to the external jitter cleaner and are externally distributed to the lpGBT transceiver banks as their transmit reference clock. (To up to 24 lpGBT links.). The transceivers then generate a 320 MHz parallel interface clock to write data into the link. As we have new trigger data in a rate of the LHC clock, we have to generate a 40 MHz Data Write Enable bit in this lpGBT link parallel clock domain to qualify which edges of the 320 MHz parallel interface clock should write data into the link. This periodic Data Write Enable bit can have 8 different phase positions relative to the 40 MHz LHC clock in the 320 MHz parallel clock domain.
4. Thus, for the deterministic transmission (writing) of the trigger data into the lpGBT links, we first have to align the (40 MHz) Data Write Enable bit of the 320 MHz clock domain with our 40 MHz internal clock generated in Point 1. Here again we sample this periodic Data Write Enable bit with our 40 MHz internal clock and if it is “1” then an alignment is found. This happens in 1 of the 8 possible phase relations, in the rest of the cases the Write Enable Bit is sampled as “0”. If it is “0”, then we shift the Data Write Enable bit with one 320 MHz clock period, and compare again. This is controlled by the lpGBT Write Enable Control state machine and the desired phase relation can be reached in a maximum of 8 steps.

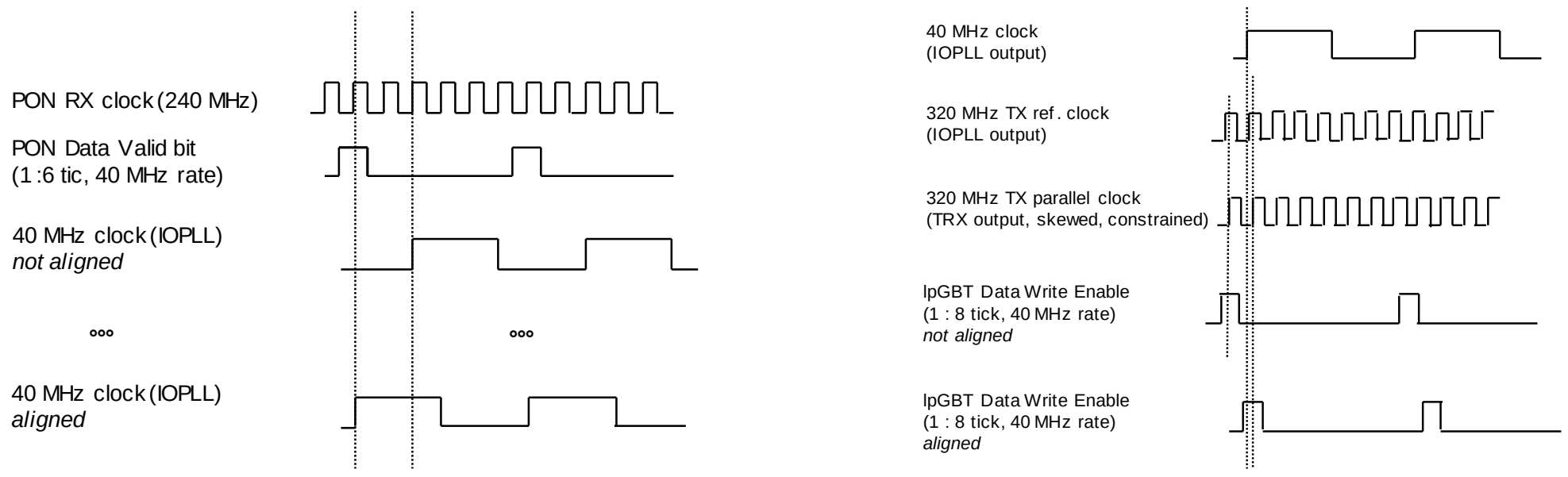

**Figure 3.** Aligning lpGBT Data Write Enable to the 40 MHz Clock

Figure 4 shows the details of the lpGBT clock and trigger distribution. The PON 240 MHz after the jitter cleaning is fed to an internal IOPLL which produces the 40 and 320 MHz. The CRU hardware has two separate external jitter cleaners. One is used for cleaning the 240 MHz, and the second one can be used to clean the lpGBT reference clock and deliver it to the FPGAs transceiver banks to support up to 24 lpGBT links.

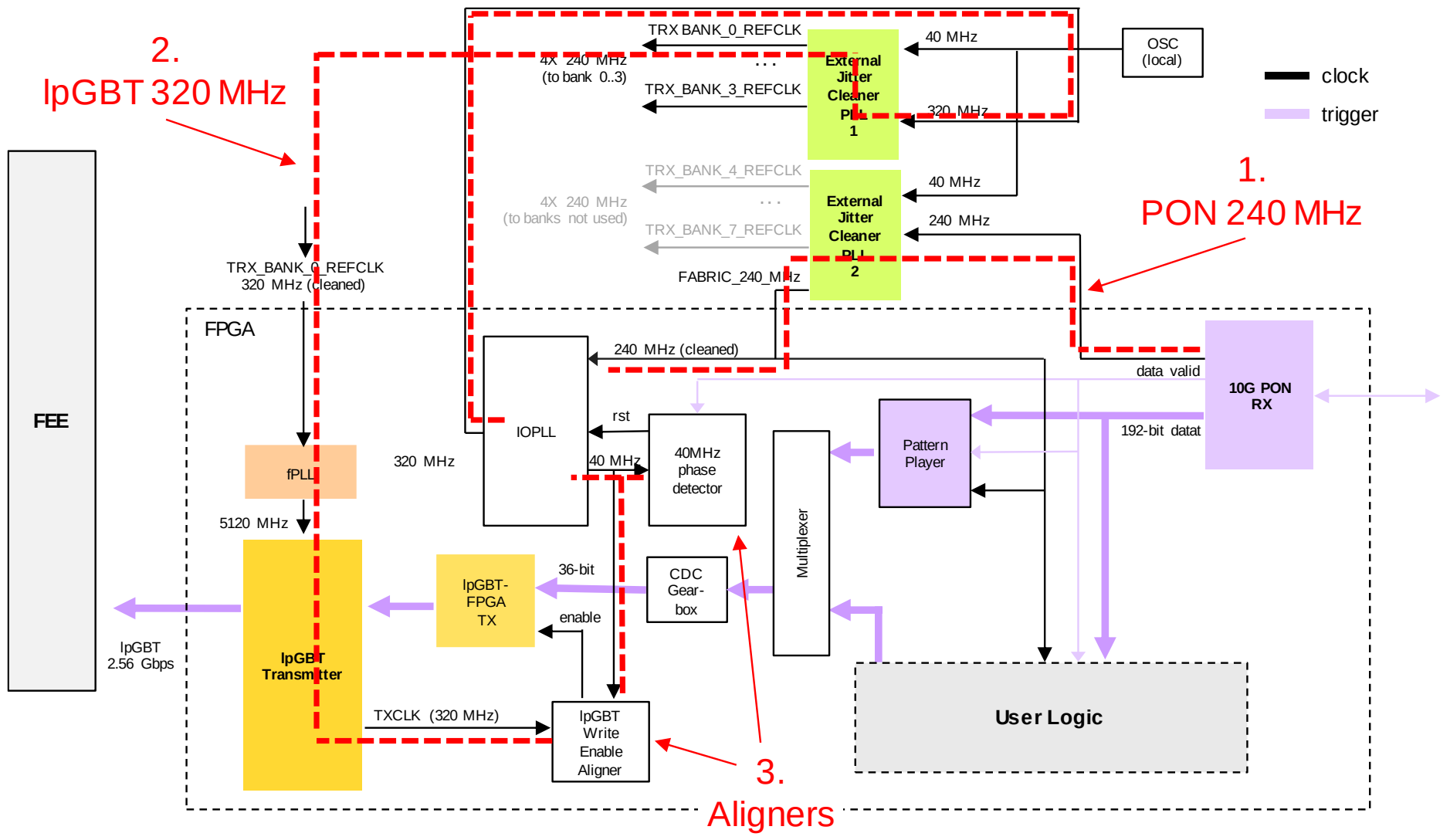

**Figure 4.** Overview of the lpGBT based clock and trigger distribution in the CRU firmware.

## 5. Perfomance tests of the lpGBT implementation

Figure 5 shows the test setup. The setup is organized around a server computer hosting one CRU with installed ALICE O2 support software. The clock and trigger information is provided by the Local Trigger Unit (LTU). The lpGBT links were tested with a VLDB+ evaluation board, hosting a VL+ optical transceiver and the lpGBT ASIC.

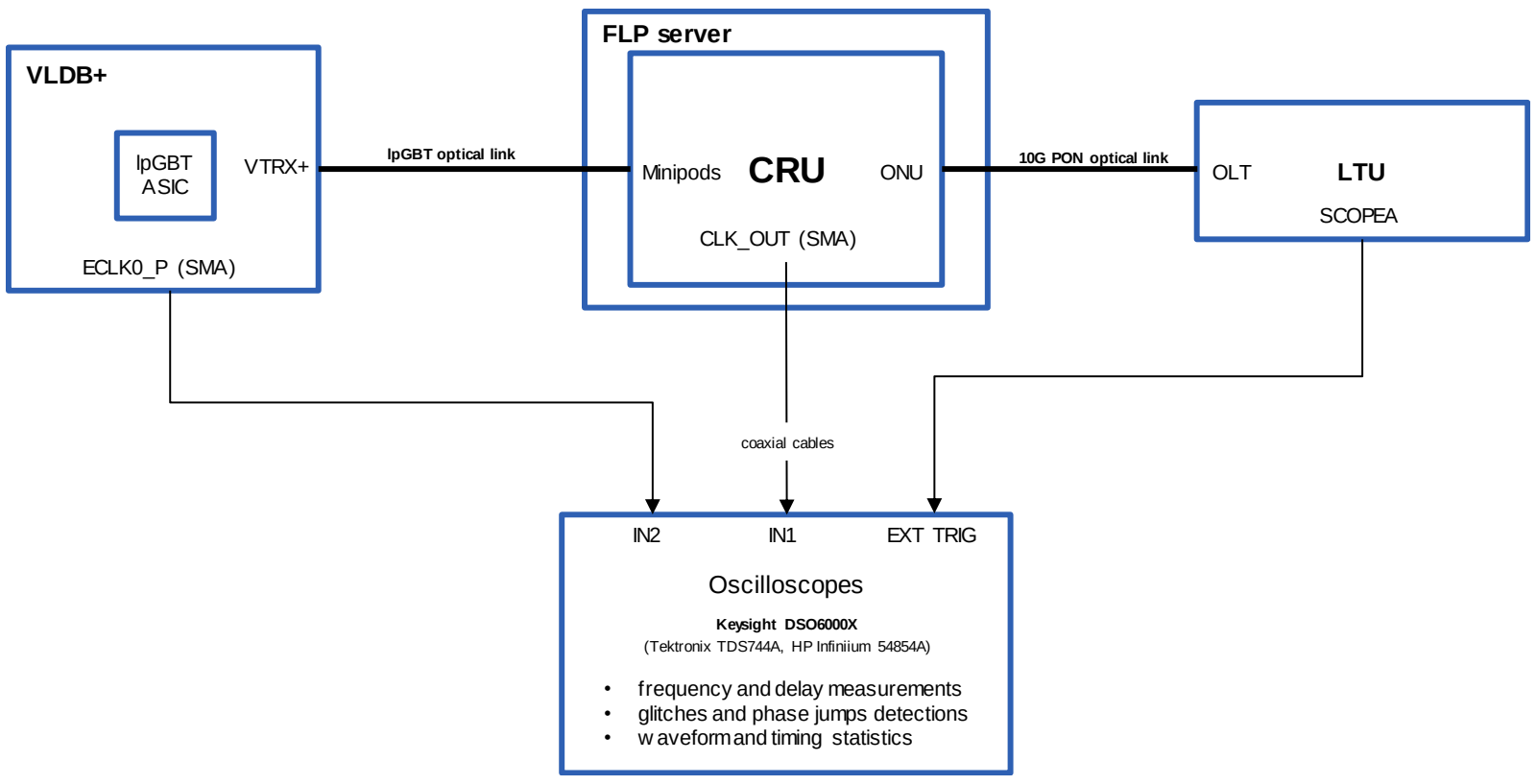

**Figure 5.** Overview of the test setup.

The implemented solution has been tested with the following test results:

1. No visible phase jumps were observed in recovered 40 MHz clock on the VLDB+ board through ~$10^{12}$ clock cycles).
2. A stable data loopback through the lpGBT ASIC in "Loopback Downlink Group Data Source" mode has been established via each of 12 lpGBT link interface of the CRU. The loopback links have been tested with a known data pattern for about 1 hour per link. (~ $10^{14}$ bits per link, without an error.)

## 6. Conclusion and Further Works

The presented solution has been successfully tested in 12 lanes with 10.24 Gb/s link speed. The downlink clock and trigger chain has fully been implemented from the ALICE LTUs down to the FEE (VLDB+), and the clock stability and deterministic latency has been successfully tested. The uplink data path has partially been integrated with the CRU firmware and software tools, and data integrity at the CRU User Logic interface has been tested. The full datapath and slow control integration is application specific and is left to the different detector projects.

The further work will concentrate on the characterization of the clock and trigger delivery in more detail (like jitter analysis, skew between links, etc.).

## Acknowledgments

This work was supported by OTKA K-135515 and NKFI 2021-4.1.2-NEMZ_KI-2022-00009 grants.